\documentclass{revtex4}

\usepackage{graphicx,epsfig}
\newcommand{\imag}{\Im {\rm m}}
\newcommand{\real}{\Re {\rm e}}
\begin{document}
\bibliographystyle{revtex}

\preprint{P3-23}

\title{Complete Reconstruction of the Neutralino System }



\author{Jan Kalinowski}
\thanks{Jan.Kalinowski@fuw.edu.pl; Preprint number: IFT-01-33}
\affiliation{Instytut Fizyki Teoretycznej,  Uniwersytet Warszawski, Warsaw, Poland }
\author{Gudrid Moortgat-Pick}
\thanks{gudrid@mail.desy.de}
\affiliation{Deutsches Elektronen--Synchrotron, DESY, Hamburg, Germany}


\date{\today}

\begin{abstract}
Consecutively to the chargino system, $\tilde{\chi}^\pm$, in which the SU(2)
  gaugino parameter $M_2$, the higg\-sino mass parameter $\mu$ and
  $\tan\beta$ can be determined, the remaining fundamental
  supersymmetry parameter, the U(1) gaugino mass $M_1$ can be analysed
  in the neutralino system, $\tilde{\chi}^0$, including its modulus and 
phase in CP--noninvariant theories. 
First experimental hints on CP violation can be 
seen in the threshold behaviour of neutralino production.
\end{abstract}

\maketitle

\vspace{-1cm}
\section{Introduction}
\label{sec:intro}
\vspace{-.4cm}
In the minimal  supersymmetric extension of the Standard Model (MSSM),
the  spin-1/2 partners of the neutral gauge 
bosons, $\tilde{B}$ and $\tilde{W}^3$, 
and of the neutral Higgs bosons, $\tilde H_1^0$ and 
$\tilde H_2^0$, mix to form neutralino mass eigenstates $\chi_i^0$
($i$=1,2,3,4). The neutralino mass matrix in the 
$(\tilde{B},\tilde{W}^3,\tilde{H}^0_1,\tilde{H}^0_2)$ basis 
{\small
\begin{eqnarray}
{\cal M}=\left(\begin{array}{cccc}
  M_1       &      0          &  -m_Z c_\beta s_W  & m_Z s_\beta s_W \\[2mm]
   0        &     M_2         &   m_Z c_\beta c_W  & -m_Z s_\beta c_W\\[2mm]
-m_Z c_\beta s_W & m_Z c_\beta c_W &       0       &     -\mu        \\[2mm]
 m_Z s_\beta s_W &-m_Z s_\beta c_W &     -\mu      &       0
                  \end{array}\right)\
\label{eq:massmatrix}
\end{eqnarray}
}
is built up by the fundamental supersymmetry parameters: the U(1) and
SU(2) gaugino masses $M_1$ and $M_2$, the higgsino mass parameter
$\mu$, and the ratio $\tan\beta=v_2/v_1$ of the vacuum expectation
values of the two neutral Higgs fields which break the electroweak
symmetry. Here, $s_\beta =\sin\beta$, $c_\beta=\cos\beta$ and
$s_W,c_W$ are the sine and cosine of the electroweak mixing angle
$\theta_W$.  In CP--noninvariant theories, the mass parameters are
complex. Without loss of generality $M_2$
can be taken real and positive so that the
two remaining non--trivial phases may be attributed to $M_1$ and $\mu$. 
Neutralinos are produced in $e^+e^-$ collisions, either in diagonal or in
mixed pairs:
$e^+ e^- \ \rightarrow \ \tilde{\chi}^0_i \ \tilde{\chi}^0_j$ 
$(i,j=1,2,3,4)$.
If the collider energy is sufficient to produce the four neutralino
states in pairs, the underlying fundamental SUSY parameters
$\{|M_1|, M_2,\,|\mu|, \Phi_1, \Phi_{\mu}; \tan\beta\}$ can be
extracted from the masses $m_{\tilde{\chi}^0_i}$ ($i$=1,2,3,4) and
couplings. Partial information from the lowest $m_{\tilde{\chi}^0_i}$
($i$=1,2) neutralino states is sufficient to extract $\{|M_1|,
\Phi_1\}$ if the other parameters have been pre--determined in the
chargino sector \cite{kneur,CKMZ}. 

\vspace{-.4cm}
\section{Mixing formalism}
\vspace{-.4cm}
In the MSSM, the four neutralinos $\tilde{\chi}^0_{1,2,3,4}$ are
mixtures of the neutral U(1) and SU(2) gauginos and the SU(2) higgsinos.
Since the matrix ${\cal M}$ is symmetric, one 
unitary matrix $N$ is sufficient to rotate the gauge eigenstate
basis $(\tilde{B}^0,\tilde{W}^3,\tilde{H}^0_1,\tilde{H}^0_2)$ 
to the mass eigenstate basis of the Majorana fields $\tilde{\chi}^0_i$:
%
${\cal M}_{diag}=N^* {\cal M} N^{\dagger}$
with
$(\tilde{\chi}^0_1, \tilde{\chi}^0_2, \tilde{\chi}^0_3, \tilde{\chi}^0_4)= 
     N (\tilde{B}, \tilde{W}^3, \tilde{H}^0_1, \tilde{H}^0_2)$.
The mass eigenvalues $m_i$ in ${\cal M}_{diag}$ can be chosen
positive by a suitable definition of the unitary matrix $N$, which
can be parametrized by 6 angles and 10 phases.
It is convenient to factorize the matrix 
$N$ into
a diagonal Majorana ${\sf M}$ and a Dirac--type ${\sf D}$ component in the  
following way:
\begin{equation}
N={\sf M}\,{\sf D}, \quad\rm{with}\quad
{\sf M}= {\sf diag}\left\{{\rm e}^{i\alpha_1},\, 
                   {\rm e}^{i\alpha_2},\,
                   {\rm e}^{i\alpha_3},\,
                   {\rm e}^{i\alpha_4}\,\right\} \label{eq:Mdef} 
\end{equation}
One overall Majorana phase is
unphysical and may be chosen to vanish and 
the matrix ${\sf D}$ can be written as a sequence of 6 two-dimensional 
rotations. Complete analytical results are given in \cite{CKMZ}. 
\vspace{-.4cm}
\section{The neutralino quadrangles}
\vspace{-.4cm}
The unitarity constraints on the elements of the mixing matrix $N$
can be formulated by means of
unitarity quadrangles which are built up by the links $N_{ik}N^*_{jk}$  
connecting two rows $i$ and $j$,  
\begin{equation}
M_{ij}= N_{i1}N^*_{j1}+N_{i2} N^*_{j2} + N_{i3} N^*_{j3}
               +N_{i4}N^*_{j4}=0 \qquad {\rm for} \ \ i\neq j
 \label{eq:M}
\end{equation}
and by the links $N_{ki}N^*_{kj}$ connecting two  columns $i$ and $j$  
\begin{equation}
D_{ij}= N_{1i}N^*_{1j}+N_{2i} N^*_{2j} + N_{3i} N^*_{3j}
               +N_{4i}N^*_{4j}=0 \qquad {\rm for} \ \ i\neq j
 \label{eq:D}
\end{equation}
of the mixing matrix.
The areas of the six quadrangles $M_{ij}$ and $D_{ij}$ are
given by
\begin{equation}
{\rm area}[M_{ij}]=\frac{1}{4}
         (|J_{ij}^{12}|+|J_{ij}^{23}|+|J_{ij}^{34}|+|J_{ij}^{41}|),
\quad
{\rm area}[D_{ij}]=\frac{1}{4}
         (|J_{12}^{ij}|+|J_{23}^{ij}|+|J_{34}^{ij}|+|J_{41}^{ij}|),
\label{eq:aD}
\end{equation}
where $J_{ij}^{kl}$ are the Jarlskog--type CP--odd ``plaquettes'': 
$J_{ij}^{kl}=\imag N_{ik}N_{jl}N_{jk}^*N_{il}^*$. The matrix
$N$ is CP violating, if either any one of the plaquettes is non--zero, or,
if the plaquettes all vanish, at least one of the
links is non--parallel to the real
or to the imaginary axis. 
The phases of the neutralino fields are fixed (modulo a common
phase) and the orientation of the neutralino quadrangles  $M_{ij}$ and
$D_{ij}$  in the complex plane is physically non--trivial. 
Since the neutralino mass matrix involves only two invariant phases $\Phi_1$
and $\Phi_{\mu}$, all the physical phases of $N$ are fully determined by these
two phases in the mass matrix. Therefore 
the experimental reconstruction of the unitarity 
quadrangles overconstrains the neutralino system and numerous consistency 
relations can be exploited to scrutinize the validity of the underlying theory
\cite{CKMZ}.
\vspace{-.4cm}
\section{Threshold behavior of neutralino production} 
\vspace{-.4cm}
In CP--invariant theories the S-wave excitation
giving rise to a steep rise $\sim \lambda^{1/2}$ of the cross section
for the nondiagonal pair near threshold
signals opposite CP parities of produced neutralinos. Not all
nondiagonal pairs of neutralinos can be produced in the S--wave; if
$\{ij\}$ and $\{ik\}$ have opposite CP parities, the pair $\{jk\}$ has
the same and it will be excited in the  P-wave characterized by the
slow rise $\sim\lambda^{3/2}$ of the cross section. 
 It is important to realize that CP-violation may change the threshold
behaviour, cf.~Fig.\ref{fig:th}a, and in particular may allow S-wave excitations
in any nondiagonal pair, e.g. finding the   $\{ij\}$, $\{ik\}$ and
$\{jk\}$ pairs to be excited in the S--wave would uniquely
signal CP violation.   
\vspace{-.4cm}
\section{Extracting the fundamental parameters}
\vspace{-.4cm}
The measurements of the chargino--pair production processes
$e^+e^-\rightarrow\tilde{\chi}^+_i\tilde{\chi}^-_j$ ($i,j$=1,2)
carried out with polarized beams can be used for a complete
determination of the basic SUSY parameters 
$\{M_2,\,|\mu|,\, \Phi_\mu\, ;\tan\beta\}$ in the chargino sector with 
high precision \cite{snow_c}.

\begin{figure}[tb]
\hskip -1cm
 \epsfig{file=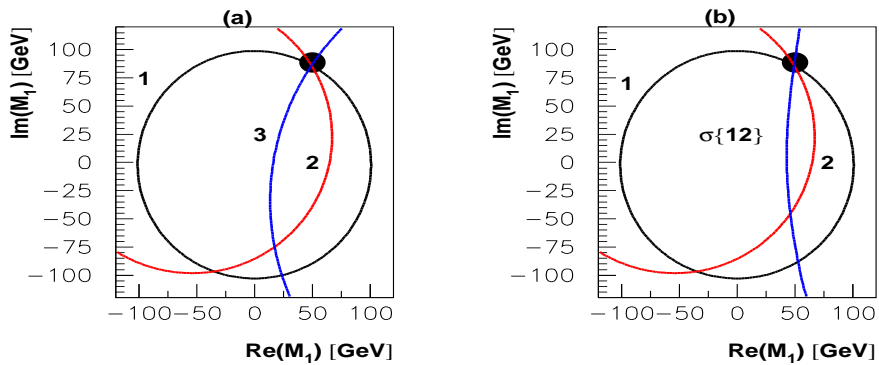, width=14cm, height=6cm}
\vspace*{-.5cm}
\caption{\it The contours of (a) three measured neutralino masses 
$m_{\tilde{\chi}^0_i}$ ($i=1,2,3$), and
(b) two neutralino masses (1,2) and one neutralino production cross section
$\sigma_{tot}\{12\}$ in the $\{\real{M_1},\, \imag{M_1}\}$ plane;
the parameter set ${\sf RP1''}$ $\{\tan\beta=10, \, M_2=190.8\, {\rm GeV},\,
|\mu|=365.1\,{\rm GeV}, \, \Phi_\mu=\pi/4\}$ is taken from the
chargino sector.}
\label{fig:contour}
\end{figure}

Each of the four characteristic equations  
for the  neutralino mass squared  
can be cast  into the form \cite{CKMZ}
\begin{eqnarray}
(\real{M_1})^2+(\imag{M_1})^2+ u_i\real{M_1}+ v_i\imag{M_1}
 = w_i \qquad (i=1,2,3,4)
\label{eq:Mphase}
\end{eqnarray}
The coefficients $u_i$, $v_i$ and $w_i$ are functions of the 
parameters $\tan\beta$, $M_2$, $|\mu|$, $\Phi_\mu$ pre--determined in
the chargino sector, and the mass
$m^2_{\tilde{\chi}^0_i}$;  the coefficient $v_i$ is necessarily
proportional to $\sin\Phi_\mu$ because physical masses are CP--even.
Each neutralino mass defines a circle in the $\{\real{M_1},\imag{M_1}\}$ plane.
Thus the measurement of three neutralino masses leads
to an unambiguous determination $M_1$,
cf. Fig.~\ref{fig:contour}a.   With only two
light neutralino masses, the two--fold ambiguity can be resolved 
by exploiting the measured cross section $\sigma\{12\}$, as shown in
Fig.~\ref{fig:contour}b.  However, if the phase $\sin\Phi_\mu$ vanishes,
there remains a two--fold discrete sign ambiguity \cite{CKMZ}.

If all four masses are experimentally accessible
the complete reconstruction of the mass and mixing parameters is easy.
The four--state mixing of neutralinos in the MSSM 
is reflected in the sum rules for the neutralino couplings \cite{CKMZ}. 
Therefore
evaluating these sum rules experimentally, it can be tested whether
the four--neutralino system $\{\tilde{\chi}_1^0, \tilde{\chi}_2^0,
\tilde{\chi}_3^0,\tilde{\chi}_4^0\}$ forms a closed system, or whether
additional states at high mass scales mix in, signaling the existence
of an extended gaugino system.
\vspace{-.4cm}
\section{The supersymmetric Yukawa couplings}
\begin{center}
\begin{figure}[htb]
\vspace*{-.8cm}
\hspace*{-.8cm}
 \epsfxsize=5.0cm \epsfbox{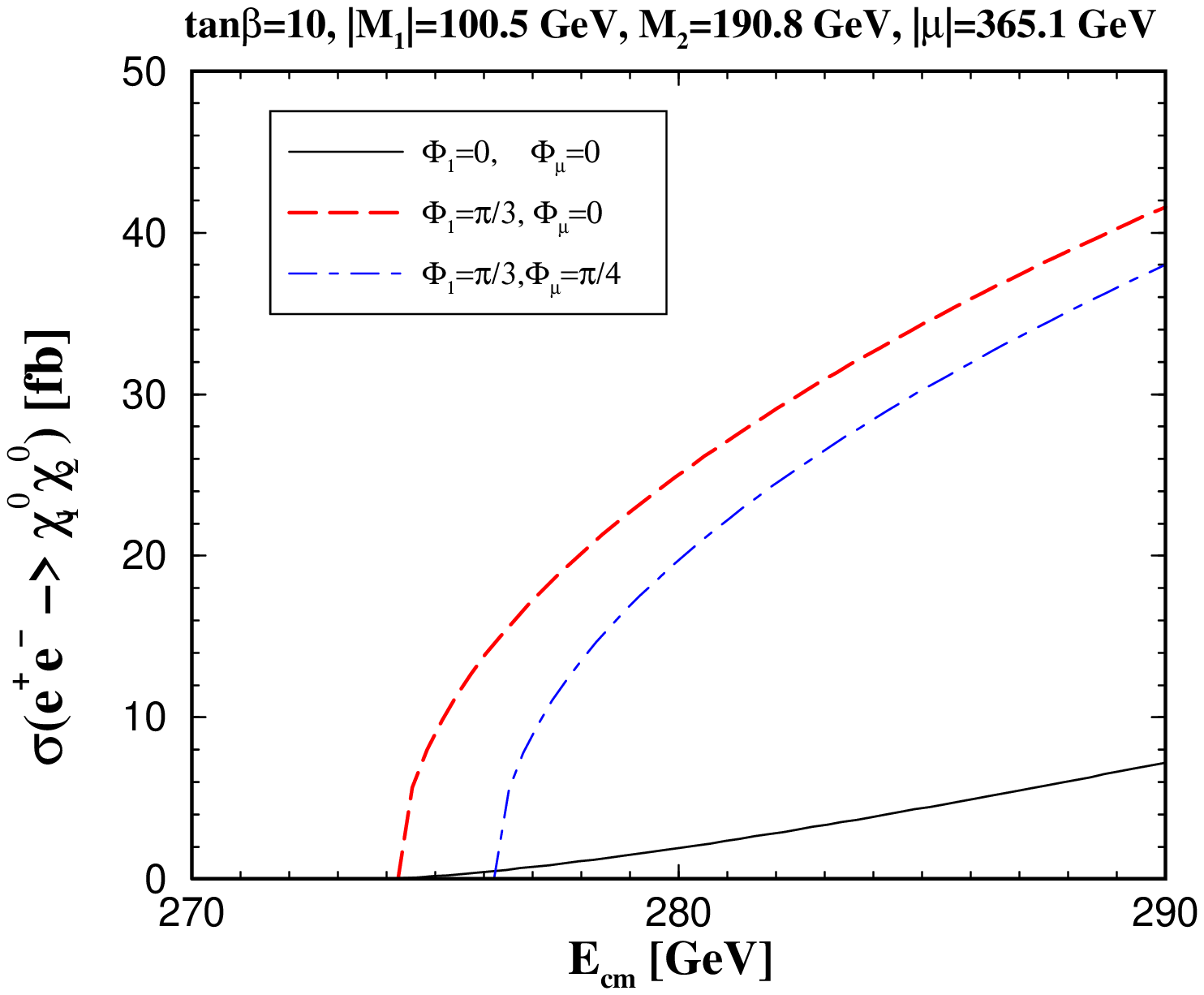}
\hspace{1.3cm}
\epsfxsize=4.7cm \epsfbox{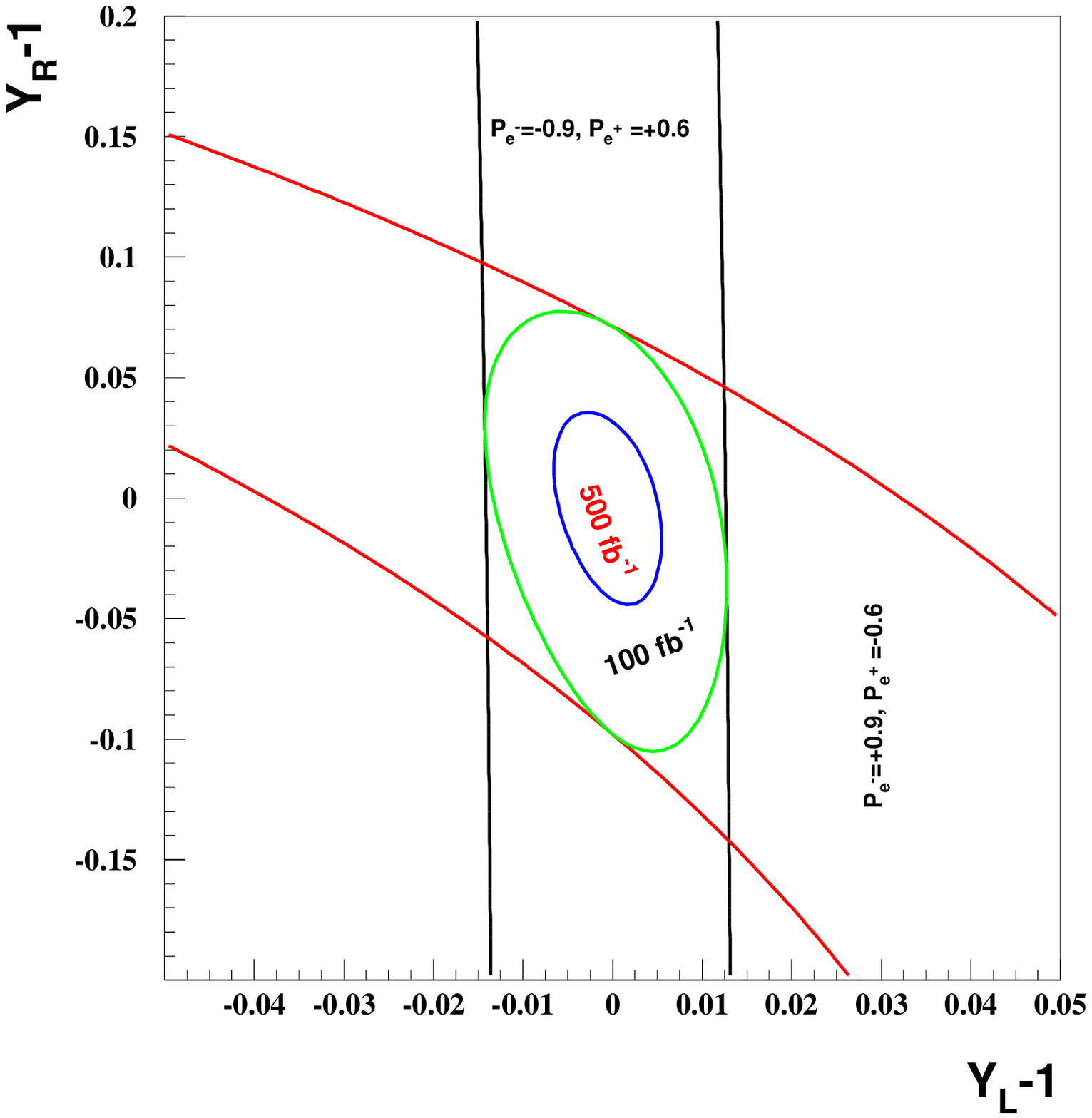}
\vspace*{-.4cm}
\caption{\it a) The threshold behaviour of the neutralino  production
  cross section $\sigma\{12\}$; the shift of the energy threshold is
  due to the dependence of the neutralino masses on the phases.\\
b) Contours of the cross sections $\sigma_L\{12\}$ and 
       $\sigma_R\{12\}$ in the plane of the Yukawa couplings $g_{_{\tilde{W}}}$
       and $g_{_{\!\tilde{B}}}$ normalized to the SU(2) and U(1) gauge
       couplings $g$ and $g'$ $\{Y_L=g_{{\tilde W}}/g,\, 
       Y_R=g_{{\tilde B}}/g'\,\}$ for the set ${\sf RP1}$ at the $e^+e^-$ 
       c.m. energy of 500 GeV; the contours correspond to the integrated 
       luminosities 100 and 500 fb$^{-1}$ and the longitudinal polarization 
       of electron and positron beams of 90\% and 60\%, respectively.}
\label{fig:th}
\end{figure}
\end{center}

\vspace{-.4cm}
A linear collider with polarized beams \cite{TDR} offers the possibility
to verify very accurately 
the fundamental SUSY assumption that the Yukawa couplings, 
$g_{_{\tilde{W}}}$ and $g_{_{\!\tilde{B}}}$ are indentical to the SU(2) and 
U(1) gauge couplings $g$ and $g'$. 
Varying the left--handed and right--handed Yukawa
couplings leads to a significant change in the corresponding left--handed
and right--handed production cross sections. Combining the
measurements of $\sigma_R$ and $\sigma_L$ for the process
$e^+e^-\rightarrow \tilde{\chi}^0_1\, \tilde{\chi}^0_2$, the
Yukawa couplings $g_{_{\tilde{W}}}$ and $g_{_{\!\tilde{B}}}$ can be
determined to quite a high precision as demonstrated in Fig.~\ref{fig:th}b.
The $1\sigma$ statistical errors have been derived
for an  integrated luminosity of 
$\int {\cal L}\, dt =100$ and $500$ fb$^{-1}$ and for 
partially polarized beams. 

\vspace{-1.5cm}
\section{Summary}
\vspace{-.4cm}
 The measurement of the processes
$e^+e^-\rightarrow \tilde{\chi}^0_i \tilde{\chi}^0_j$ ($i,j$=1,2,3,4),
carried out with polarized beams and combined with the analysis of the
chargino system $e^+e^-\rightarrow \tilde{\chi}^+_i \tilde{\chi}^-_j$
($i,j$=1,2), can be used to perform a complete and precise analysis of
the basic 
SUSY parameters in the gaugino/higgsino sector $\{M_1, M_2, \mu;
\tan\beta\}$. The closure of the neutralino system can be verified by 
exploiting the sum rules for  production cross sections.

\vspace{.3cm}
JK was supported in part by the KBN Grant 5 P03B 119 20 (2001-2002) and the 
European Commision 5-th framework contract HPRN-CT-2000-00149. GMP was 
partially supported by the DPF/Snowmass Travel Fellowship from the
Division of Particles and Fields of the American
Physical Society, and of the Snowmass 2001 Organizing Committee.
The authors would like to thank S.Y. Choi and P.M. Zerwas for many lively
and interesting discussions.
\vspace{-.5cm}


\end{document}